\documentclass[12pt]{article}

\begin{document}

\newcommand{\be}{\begin{equation}}
\newcommand{\ee}{\end{equation}}
\newcommand{\bea}{\begin{eqnarray}}
\newcommand{\eea}{\end{eqnarray}}
\newcommand{\non}{\nonumber}
\begin{titlepage}
\begin{center}
{\Large\bf Modified Dispersion Relations\\ 
\vspace*{0.2cm}
and trans-Planckian Physics}\\
\vspace*{1cm} Massimiliano Rinaldi\footnote{E-mail:
rinaldim@bo.infn.it} \\
\vspace*{.5cm}
{\em Dipartimento di Fisica dell'Universit\`a di Bologna,\\
Via Irnerio 46, 40126 Bologna, Italy.}
\vspace*{1cm}

\begin{abstract}
\noindent We consider modified dispersion relations in quantum field theory on curved space-time. Such relations, despite breaking the local Lorentz invariance at high energy, are considered in several phenomenological approaches to quantum gravity. Their existence involves a modification of the formalism of quantum field theory, starting from the problem of finding the scalar Green's functions up to the renormalization of various quantum expectation values. In this work we consider a simple example of such modifications, in the case of ultra-static metric. We show how to overcome the lack of Lorentz invariance by introducing a preferred frame, with respect to which we can express the Green's functions as an integral over all frequencies of a space-dependent function. The latter can be expanded in momentum space, and by integrating over all frequencies, we finally find the expansion of the Green's function up to four derivatives of the metric tensor. The relation with the proper-time formalism is also discussed.
\end{abstract}

\end{center}
\end{titlepage}

\section{Introduction}

Modified dispersion relations (MDRs) have been recently considered in various models of high-energy physics. Indeed, MDRs can be used as a phenomenological approach to investigate physics at the Planck scale, where General Relativity is no longer reliable. In this sense, MDRs can be seen as a phenomenological approach to shed some light on the underlying quantum theory of gravity. The main idea is that, at the Planck scale, one or more postulates of quantum field theory are no longer true, and local Lorentz invariance is the first candidate. In other words, Lorentz invariance is considered as an approximate symmetry, valid only in the low (with respect to the Planck value) energy regime.
However, as the Planck scale is very small ($10^{-35}\,m$), it is natural to ask whether trans-Planckian effects are really important. In fact, there are at least two physical phenomena, where they might lead to macroscopic effects.
The first is certainly in cosmology. Several authors believe that trans-Planckian effects affected the early-stage evolution of the Universe, and that they left some observable fingerprints, e.g. in the CMB inhomogeneities (see, for example, \cite{MDRcosmo}). The second concerns black hole physics, where the quantum thermal emission discovered by Hawking \cite{HAW} is related to modes of arbitrarily large frequency near the horizon. Besides these specific situations, there are also some fundamental theories, such as String Theory, Loop Quantum Gravity, and Double Special Relativity, which predict MDRs (see the review \cite{mattingly} for references). 

Some people believe that, even in extremely high energy conditions, trans-Planckian physics is not relevant. For example, Starobinsky has shown that trans-Plankian physics does not affect inflation \cite{staro}. In the case of black holes instead, Unruh et al.\ have proven that the spectrum emitted at infinite distance from the hole is only marginally affected by MDRs \cite{Unruh1}. However, these ideas are based on the assumption that modified high energy modes do not affect sensibly the quantum back-reaction on the background. In the semi-classical approach to field theory on curved space \cite{BirDav}, the classical energy-momentum tensor is replaced by its quantum expectation value, according to
\bea
    R_{\mu\nu}-{1\over 2}Rg_{\mu\nu}=8\pi G\langle T_{\mu\nu}\rangle~.
\eea
This quantity, which is not vanishing in general, even in vacuum, acts as a source, which can modify the geometry of space-time (back-reaction). Therefore, if its value becomes large because of MDRs, then physics is certainly affected by modified dispersion relations. 

It is well known that $\langle T_{\mu\nu}\rangle$ is in general a diverging quantity, even in flat space, and it needs to be renormalized. If MDRs are present, usual renormalization techniques are no longer reliable. Some results have been achieved in the context of cosmology in \cite{Mazzitelli2}. However, in the case of non-homogeneous backgrounds, the renormalization of the stress tensor in the presence of MDRs, appears much more difficult. The Klein-Gordon operator, unlike the cosmological case, now contains spatial derivatives of (at least) fourth order  \cite{Corley}, thus the usual deWitt-Schwinger representation of the Green's functions does not work.

In this paper, we  consider the problem of MDRs on non-homogeneous manifolds by first looking at the ultra-static case. Our method relies upon the existence of a preferred frame associated to a unit time-like vector field $u^{\mu}$ \cite{Jacobson1}-\cite{LLMU}. If the space-time is ultra-static, the Green's functions can be written as an integral over all frequencies of a function, which is independent of the time associated to the observer co-moving with $u^{\mu}$. This function satisfies an equation, which can be solved in momentum space by applying the well-known Bunch and Parker procedure \cite{BP}. Thus, we can find a momentum-space representation of the time-independent part of the Green's functions.
 
The plan of the paper is the following: in the next section, we introduce the modified dispersion relations and discuss the preferred frame. In section 3 we perform the deWitt-Schwinger analysis, and in section 4 we find the expansion of the Green's function in momentum space and in coordinate space. Throughout this paper, we use the signature $(-,+,+,+\cdots)$, and set $\hbar=c=1$

\section{Modified dispersion relations}

\noindent In Minkowski space-time the dispersion
relation for a  scalar field of mass $m$ can be found by inserting the ansatz $\phi \sim
\exp(-ik_0t+i\vec k\cdot \vec x)$ in the Klein-Gordon
equation $(-\nabla^2+m^2)\phi=0$, which, together with the identity $\omega_k=|k_0|$, yields $\omega_k^2=|\vec k|^2+m^2$. A general
dispersion relation can be written as $ \omega_k^2=|\vec k|^2+{\cal F}(|\vec k|)+m^2$, where ${\cal F}(|\vec k|)$ is a scalar
function of the modulus of the wave-vector $\vec k$. If ${\cal F}$ depends on the square of the modulus, rotation invariance is preserved. If it is also
analytic, then it can be expanded, and, to leading order, the MDR reads \bea\label{MDR} \omega_k^2=|\vec k|^2+\epsilon^2
|\vec k|^4+m^2~, \eea where $\epsilon$ is a cut-off parameter that
sets the lowest value of $\vec k$ at which corrections to the
dispersion relation are ignored. Also, the sign of $\epsilon^2$
indicates wether the modes are sub-luminal ($\epsilon^2 <0$) or
super-luminal ($\epsilon^2>0$).

Despite MDRs such as (\ref{MDR})  break Lorentz invariance, general covariance can be preserved if there is a  preferred frame associated to a unit time-like vector, as shown by Jacobson et al. \cite{Jacobson1}. The action considered in this work is given by
\begin{equation}
S=\int
d^4x\sqrt{-g}\left[-a_1R-b_1F^{ab}F_{ab}+\lambda(g_{ab}u^au^b+1)\right]~,
\end{equation}
where $a_1$ and $b_1$ are constant parameter,
$F_{ab}=2\nabla_{[a}u_{b]}$, and $\lambda$ is a Lagrange multiplier
that ensures $u^a$ to be a unit time-like vector. One can also construct a massless scalar field Lagrangian, which preserves general covariance, such that the Klein-Gordon equation reads \cite{Jacobson1}
\bea\label{modklein}
(\nabla^2-m^2-\epsilon^2\hat\nabla^4)\,\phi(x)=0~.\eea
The operator $\hat\nabla^4$ is the square of the
covariant spatial Laplacian  \bea\label{spacelap}
\hat\nabla^2\phi=-q^{ac}\nabla_{a}(q_c^{\,\,\,b}\nabla_b\phi)~,\eea where $q_{ab}$ is the projector operator on the spatial sections, defined
by $ q_{ab}=g_{ab}+u_au_b$. 
In order to evaluate the effects of MDRs at the level of back-reaction, we first need to compute the renormalized Green's functions. In order to do so, we begin by evaluating the deWitt-Schwinger expansion, which leads to a momentum-space representation for the Green's functions.

\section{The deWitt-Schwinger expansion}

In this section we consider the deWitt-Schwinger construction \cite{dewitt}-\cite{fulling}, suitably modified by MDRs \cite{ME}. We work with a $(n+1)$-dimensional  globally hyperbolic manifold, such that it can be foliated into space-like surfaces of constant $\tau$ \cite{LLMU}. The parameter $\tau$ can be used to define the unit time-like tangent vector $u_{\mu}=-\partial_{\mu}\tau$, with respect to some coordinate system $x^{\mu}$. Therefore, $\tau$ is the time relative to a free-falling observer moving with velocity $u^{\mu}$. Thus, the metric assumes the form \bea\label{Jmetric}
ds^2=g_{\mu\nu}dx^{\mu}dx^{\nu}=-d\tau^2+q_{\mu\nu}dx^{\mu}dx^{\nu}~.\eea where $d\tau=u_{\mu}dx^{\mu}$. This form guarantees that $\det g_{\mu\nu}= \det q_{ij}$ where Latin indices label spatial coordinates only.  In this paper, we consider the case when $q_{\mu\nu}=g_{\mu\nu}+u_{\mu}u_{\nu}$ does not depend on $\tau$, i.e. when the metric is ultra-static. Hence, the d'Alambertian operator splits as $\nabla^2\phi=\hat\nabla^2\phi-\ddot\phi$, where the dot indicates derivative with respect to $\tau$.

We now consider the modified Klein-Gordon equation (\ref{modklein}),
 for which the associate Green's functions satisfies the equation
\begin{equation}
(\nabla^2-m^2-\epsilon^2\hat\nabla^4)\,{\cal G}(x,x')=-g^{-1/2}\delta^{(n+1)}(x-x')~.
\end{equation} Here, and from now on, we denote $g(q)\equiv \det | g(q)|$. Because of the MDR, the function ${\cal G}$ is no longer Lorentz invariant, but, since the modification is a quartic spatial operator, the $O(n)$ rotational invariance holds. It is therefore convenient to express the Green's functions as 
\bea\label{greensep}
{\cal G}(x,x')=\int\frac {d\omega}{ 2\pi}\,e^{i\omega(\tau-\tau')}\,G(x^j,x'^j,\omega)~,
\eea where $G(x^j,x'^j,\omega)$ depends on spatial coordinates and $\omega$ only. Thus, one can show that $G$ satisfies the $n$-dimensional equation
\bea\label{greeen}
(\hat\nabla^2-m^2+\omega^2-\epsilon^2\hat\nabla^4)\, G(x^j,x'^j,\omega)
=-q^{-1/2}\delta^{(n)}(x^j-x'^j)~.\label{spaceGreen}
\eea

\section{Momentum space representation}

To solve Eq.~(\ref{greeen}), we use the method due to Bunch and Parker \cite{BirDav},\cite{BP}, which essentially makes use of a local Taylor's expansion of the metric tensor expressed in Riemann normal coordinates \cite{Christensen},\cite{ Poisson}. To begin with, we define the function $\bar G$ such that
$ G(x,x',\omega)=q^{-1/4}(x)\bar G(x,x',\omega)q^{-1/4}(x')$ and write $
q^{-1/2}(x)\delta(x-x')=q^{-1/4}(x)\delta(x-x')q^{-1/4}(x')$. In this form, these functions behave as bi-scalars at $x$ and $x'$ \cite{Poisson}. Here, $x$ and $x'$ are considered as separate point on the \emph{same spatial slice} $\tau=$ const. Next, we introduce the Riemann normal coordinates $y$ with origin at $x'$. Thus, $q(x')=1$ and the Green's function $\bar G$ satisfies \bea\label{redgreen}
q^{1/4}(\omega^2+m^2-\hat\nabla^2+\epsilon^2\hat\nabla^4)(q^{-1/4}\bar G(y))=\delta(y)~.
\eea  In a neighborhood of $x'$, we can expand the induced metric in Riemann normal coordinates as \cite{BP}
\bea
q_{mn}&=&\delta_{mn}-{1\over 3}\hat R_{manb}y^ay^b-{1\over 6}\hat R_{manb;p}y^ay^by^p+\\\non
&+&\left(-{1\over 20}\hat R_{manb;pq}+{2\over 45}\hat R_{ambl}\hat R^{\,l}_{~pnq}\right)y^ay^by^py^q~,
\eea
where all coefficients are evaluated at $x'$ (i.e. at $y=0$) and contain up to four derivatives of the metric $q_{ij}$. With the help of this expansion, we find
\bea
q^{1/4}\hat\nabla^2(q^{-1/4}\bar G(y))\simeq \delta^{ij}\partial_i\partial_j\bar G+{1\over 6}\hat R\bar G+{1\over 6}\hat R_{;j}y^j\bar G+\hat H_{ij}y^iy^j\bar G~,
\eea where
\bea
\hat H_{ij}=-{1\over 30}\hat R^{p}_{~i}\hat R_{pj}+{1\over 60}\hat R^{p}_{~i}{}^{q}_{~j}\hat R_{pq}+{1\over 60}\hat R^{pql}{}_{i}\hat R_{pqlj}+{3\over 40}\hat R_{;ij}+{1\over 40}\hat R_{ij;p}{}^p~.
\eea The expansion above is also obtained by using the fact that $\bar G$ depends on $y^2$, being rotationally invariant. To find the expansion for the quartic operator, we can proceed by iteration into momentum space. Indeed, if we define the local Fourier transform of $\bar G$ as
\bea\label{fourier}
\bar G(y)=\int{d^nk\over (2\pi)^n}\,e^{ik\cdot y}\tilde G(k)~,
\eea
one can show that \cite{ME}
\bea\label{psi2}\non
q^{1/4}\hat\nabla^4(q^{-1/4}\bar G)&=&k^4\tilde G-{k^2\over 3}\hat R\tilde G-{1\over 3}\hat R_jk^j(\tilde G+2k^2D\tilde G)+{1\over 36}\hat R^2\tilde G\\\non
&+&2\hat H(\tilde G+2k^2D\tilde G)+8\hat H_{ij}k^ik^j(k^2D^2\tilde G+D\tilde G)~,
\eea 
where $\hat H\equiv \delta^{ij}\hat H_{ij}~$, and $
\partial/\partial k_j=2k^j\partial/\partial |\vec{k}|^2\equiv 2k^jD$.
With these elements, we can  expand Eq.~(\ref{redgreen}) in momentum space up to four derivatives of the metric as
\bea\non
&1&=(k^2+\epsilon^2k^4+m^2-\omega^2)\tilde G+\\\non
&-&{1\over 6}\hat R\tilde G(1+2\epsilon^2 k^2)-{i\over 3}\hat R_{;j}k^j(D\tilde G+\epsilon^2\tilde G+2\epsilon^2k^2D\tilde G)+\\\non
&+&{\epsilon^2\over 36}\hat R^2\tilde G+2\hat H(D\tilde G+\epsilon^2\tilde G+2\epsilon^2k^2D\tilde G)+\\
&+&4\hat H_{ij}k^ik^j(D^2\tilde G+2\epsilon^2k^2D^2\tilde G+2\epsilon^2D\tilde G)~.
\eea 
At the zeroth order, this equation yields
\bea
\tilde G_0=\frac{1}{k^2+\epsilon^2k^4+m^2-\omega^2}~,
\eea while, as $\tilde G=\tilde G_0+\tilde G_2+\ldots$, the following orders can be found by recursion, finally yielding our main result
\bea\non
\tilde G&=&\tilde G_0-{1\over 6}\hat RD\tilde G_0-{i\over 12}\hat R_{;j}\tilde\partial^jD\tilde G_0+\left({1\over 72}\hat R^2-{1\over 3} \hat H\right)D^2\tilde G_0+\\
&-&{1\over 3}\hat H_{ij}\tilde\partial^i\tilde\partial^jD\tilde G_0~,\label{finalexp}
\eea 
where tilded derivatives are with respect to $k^i$. We conclude this section, by connecting the above expansion with the proper time formalism of deWitt-Schwinger.
At $\omega$ fixed (i.e. on a given spatial slice), we define
\bea\label{srepr}
\tilde G_0=i\int_0^{\infty}ds\, e^{-is(k^2+\epsilon^2k^4+m^2-\omega^2)}~,
\eea and replace into (\ref{finalexp}). By swapping integrals, we find \cite{ME}
\bea
\non G(y)&=&i\int_0^{\infty}ds\,e^{-is(m^2-\omega^2)}\Big[1+(is)(1-2\epsilon^2\partial^2)f_1+\\
&+&2(is)^2\epsilon^2f_2+(is)^2(1-2\epsilon^2\partial^2)^2f_2\Big]\, I_{\epsilon}(y,s).\label{dewitt}
\eea
In this expression,
\bea
f_1={1\over 6}\hat R+{1\over 36}\hat R_{;j}y^j-{1\over 3}\hat H_{ij}y^iy^j~,\quad
f_2={1\over 72}\hat R^2-{1\over 3}\hat H~,\eea
and 
\begin{equation}\label{Int}
I_{\epsilon}(y,s)={e^{{iy^2\over 4s}}\over (4is\pi)^{n/2}}\,\sum_{\lambda=0}^{\infty}{1\over \lambda !}\left(\frac{i\epsilon^2}{16s}\right)^{\lambda}{\cal H}_{[4\lambda]}\left(\frac{\vec{y}}{\sqrt{4is}}\right).
\end{equation}
We see that this  expansion in proper time yields a sum of the Hermite polynomials ${\cal H}_{[4\lambda]}$  of order $4\lambda$, and of their derivatives, which does not seem to converge to any known function. In the Lorentz invariant case instead, the sum is trivial, and the integral over $s$ converges to Hankel's functions of second kind \cite{dewitt,Christensen}. 

\section{Summary}

\noindent In this work we present our first results concerning quantum field theory on curved backgrounds, with modified dispersion relations. In particular, we carefully analyze the Klein-Gordon equation associated to a scalar field propagating on a ultra-static space-time, as a first step towards the physically relevant case of stationary metrics. By assuming the existence of a preferred frame, and despite the breaking of local Lorentz invariance, we still have general covariance and, above all, rotational invariance over slices of constant time. This fact allows for a dimensional reduction of the Klein-Gordon equation, so that one can consider unambiguously the Green's functions at a fixed frequency, as measured by the free-falling observer. Then, by following the method of Bunch and Parker, we finally obtained the expansion of the Green's functions up to four derivatives of the metric in momentum space and in coordinate space.

\vspace{\baselineskip}		
\noindent {\bf Acknowledgements:} I wish to thank R.~Balbinot, A.~Fabbri, S.~Fagnocchi, R.~Parentani, P.~Anderson, A.~Ottewill, and M.~Casals for helpful discussions.

\end{document}